\def\ket#1{| #1 \rangle}
\def\bra#1{\langle #1 |}
\def\lrp{\buildrel\leftrightarrow\over\partial}

\def\parslash{\hbox{\rm\rlap/$\partial$}}

\documentclass[preprint,prd,aps,showpacs,draft,floatfix]{revtex4}
\usepackage{supertabular}
\begin{document}
\input psfig.sty
%
%
\def\Journal#1#2#3#4{{#1} {\bf #2}, #3 (#4)}
%
%
\def\EPJC{{\em Eur. Phys. J.} C}
\def\MPLA{{\em Mod. Phys. Lett.} A}
\def\NCA{\em Nuovo Cimento}
\def\NIM{\em Nucl. Instrum. Methods}
\def\NIMA{{\em Nucl. Instrum. Methods} A}
\def\NPB{{\em Nucl. Phys.} B}
\def\PLB{{\em Phys. Lett.}  B}
\def\PRD{{\em Phys. Rev.} D}
\def\PRL{\em Phys. Rev. Lett.}
\def\PRPT{\em Phys. Rept.}
\def\SJNP{\em Sov. J. Nucl. Phys.}
\def\SP{\em Sov. Phys.}
\def\ZPC{{\em Z. Phys.} C}
%
%
\def\st{\scriptstyle}
\def\sst{\scriptscriptstyle}
\def\mco{\multicolumn}
\def\be{\begin{equation}}
\def\ee{\end{equation}}
\def\bea{\begin{eqnarray}}
\def\eea{\end{eqnarray}}
\def\CPbar{\hbox{{\rm CP}\hskip-1.80em{/}}}
%
%
\def\Bbar{\overline{B}}
\def\cbar{\overline{c}}
\def\Dbar{\overline{D}}
\def\Kbar{\overline{K}}
\def\dzbar{{\overline{D}\,^0}}
\def\dz{{D^0}}
\def\dmix{$D^0-\overline{D}^0$ mixing}

\title{On the $D^0$ -- $D_s$ lifetime difference and $\tau\to 7\pi + \nu_\tau$ decays}

\author{S. Nussinov$^{1,2}$}
\author{M. V. Purohit$^2$}

\affiliation{$^1$School of Physics and Astronomy, \\
         Tel Sackler Faculty of Exact Sciences,\\
         Tel Aviv University and \\
         $^2$Dept. of Physics \&\ Astronomy,\\
         Univ. of South Carolina, SC 29208}
\date{\today}

\begin{abstract}
  In this paper we discuss some aspects of inclusive decays of charmed
mesons and also decays of the $\tau$ lepton into $\nu_\tau + 7\pi$.  We
find that phase space effects are likely to explain the observed
lifetime ratio $\tau(D_s^+) / \tau(D^0)$ = 1.17.  In particular one need
not appeal to a large annihilation contribution in the inclusive $D^0$
decay which, being absent in $D_s^+$ decays could also contribute to the
enhanced $D^0$ decay rate relative to that of the $D_s^+$. Examining a
separate problem, we find that the rate for $\tau\to
\nu_\tau + 7\pi$ is almost completely dominated by the tiny phase space
for the final eight particle state.  Using an effective chiral
Lagrangian to estimate the matrix element yields a branching
ratio into the channel of interest far 
smaller than the present upper bound.
\end{abstract}
\pacs{13.20.Fc, 13.25.Ft, 13.35.Dx, 14.40.Lb}

\maketitle

\section{Introduction}

      Both the (spin avaraged) square of the invariant amplitude, 
$\sum |\overline{M}_{if}|^2$
and the Lorentz invariant phase space (LIPS) enter into the calculation 
of any decay or scattering process. In attempting to calculate decays of 
charmed particles we often find that (at least!) one of these factors 
is not well known. Thus for a specific decay say $D^0\to K\pi\pi$, or
$D_s^+\to KK\pi$ the
observed masses of the final and initial particles entering the LIPS 
are precisely known. On the other hand the relatively low
energies of the final hadrons $O(M/3 = 0.6\;\hbox{GeV})$, precludes any perturbative
approach to the calculation of the decay matrix ellement.

      For inclusive (semi- or non-leptonic) decays a perturbative 
approach is used. In the $\Lambda_{\hbox{QCD}}/M_Q \to 0$ and / or
$F_{\hbox{D,B}}/M_{\hbox{D,B}}\to 0$  limits we 
neglect the spectator quark and view the process as the decay of a 
``free'' heavy quark: $Q \to q\overline{l}\nu$  or 
                         $Q \to qu\overline{d}$, with $Q=c$ and $q=s$ for the CKM favored 
decays. One uses here the ``Hadron Quark Duality'' assumption that
for the inclusive decay of interest, hadronization
generates just a ``Final State Interaction Phase''.
This ``Phase'' is a unitary matrix which  shuffles decay
probabilities between different channels without modifying inclusive rates. 

      The four-quark ``Effective Lagrangian'' describing the $Q$ decay includes 
QCD normalization effects due to integration over momenta between the
high $W$ mass and the intermediate $M_Q$ scale. Unlike 
the ultimate ``Renormalization'' yielding a ``Chiral 
Effective Lagrangian'' appropriate for scales of $\Lambda_{\hbox{QCD}} \approx 300$~MeV (and which 
would then directly yield the amplitudes for specific hadronic decays), 
the effective 4-quark Lagrangian can be estimated via perturbative QCD.
The resultant enhancement of the ordinary $c\, \overline{s}\, \overline{d}\, u$ term and the 
negative interference (for the D decays) with the new
$c \overline{u} \overline{s} d$ term helps explain the reduction of 
the semileptonic branchings and the $\Gamma(D^0) / \Gamma(D^+) \approx 2$ width ratio.

The latter could in part be due to annihilation with the spectator $\overline{u}$
quark which is present in the $D^0$ but not in the $D^+$. While such annihilation transcends the
spectator model it is suppressed by a $F_q / M_q$ factor.  

      However the inclusive calculation of actual {\it rates} (rather than ratios 
thereof) includes also as essentially a phase space factor: the
$M_Q^5 f(r)$ factor with $r = m/M = 0.1 -- 0.07$. Neither the charm nor
the strange 
quark masses relevant for this decay are independently measured and 
even a 2-3\%\ variation of $M$ can cause a 10-15\%\ change in the actual decay rate.
 
      Recent experiments which measure the $D_s^+$ lifetime with 3\%\ precision suggest
that it is longer than the $D^0$ lifetime by $\sim$17\%.
This modest difference can be due, in particular, to the following two 
distinct effects:

\begin{enumerate}
\item The above mentioned annihilation amplitude contributing to
the $D^0$ width which is practically absent in $D_s$ decays due to a large
helicity $(m_u/m_c)^2$ suppression of the final multipion states.\cite{fn1}

\item Due to the helicity suppression of decays into multipions, the 
final states in $D_s^+$ decays contain a $K$ and a $\overline{K}$ rather 
than a $K$ and a $\pi$ as in the ``corresponding'' D decay, e.g., 
$D_s^+ \to K^+K^-\pi^+$ versus $D^0\to \overline{K}^0\pi^+\pi^-$. This,
in turn, may reduce the corresponding phase space. 
\end{enumerate}

      Elaborating on point 2 we note that the $D_s^+$-$D^0$ mass
difference ($= 104$ MeV), is the smallest of the mass differences
between strange and corresponding non-strange mesons and baryons:
 
\begin{eqnarray}
  m_K^+ - m_\pi^+        &=& 354.1\;\hbox{MeV}\\
  m_{K^{*+}} - m_\rho^+  &=& 122.4\;\hbox{MeV}\\
  m_\Lambda - m_n        &=& 176.1\;\hbox{MeV}
\label{eqn_DM}
\end{eqnarray}
 
      Hence phase space effects in the actual physical 
final multiparticle states, for intermediate resonances,
and even within the spectator model, reduce $\Gamma(D_s^+) / \Gamma(D^0)$. 
Conceivably this could account for the deviation of this ratio from unity.  

      In the second half of the paper we consider the decay of the tau
lepton into final states with many 
pions. This is motivated by the fact that strong upper bounds of
$1.8\times 10^{-5}$ and $3\times 10^{-6}$ have been obtained
for the branching ratio into the state with seven charged pions. We
find that this decay rate is almost completely dominated by the decrease
in phase space as the energy per final state pion is reduced. We 
use the simplest chiral Lagrangian to estimate the relevant matrix
elements and decay rate. The estimated branching
ratios are far smaller than the above upper bounds - making the
observation of these decays extremely unlikely.  We argue that despite
the existence of many resonances in the 3 and 4 pion channels the
utilization of an effective local Lagrangian approach is indeed
justified. The condition for this is that the resonances are {\bf
broad} as they are in the case of interest. We also address
possible unusual ``collective'' enhancements which could occur for a
sufficiently large number of pions. We find however that for the case at
hand with only seven pions no such enhancement is likely to occur.
 
\section{LIPS for specific hadronic decays}

      We have examined in detail the difference in decay rates of the $D^0$ and 
the $D_s^+$ due to phase space. We began by taking the entire list of
$D^0$ decay modes from the Particle Data Group (PDG)\cite{pdg}. The sum
of the $D^0$ branching ratios listed there is approximately 122\%\ where
the excess is due to quantum interference. This indicates that measurements
of the $D^0$ decay modes are substantially complete and can be used to
estimate the total $D^+_s$ rate if we assume that for each $D^0$ decay
mode there is a corresponding $D_s^+$ decay mode (see table
\ref{tbl_d0ds} below) and
that the corresponding partial widths are simply the square of a common
matrix element times the phase space of each decay. We thus compute the
expected partial widths for the $D^+_s$ decay modes corresponding to
each $D^0$ decay mode and sum to obtain the total decay rate. 

      While this procedure may not be precise for individual decay modes
due to final state interactions, 
it should correctly approximate the phase space effect when we sum over all the
decay modes. In implementing the procedure we took into account the
variation of phase space due to width of resonances. When converting
$s\overline{s}$ final states to $\eta$ and $\eta^\prime$ we used the
simplest mixing described in the PDG.\cite{pdg} Similarly, we followed
the PDG in assuming that the $K_{1A}$ is an equal mixture of $K_1(1270)$
and $K_1(1400)$. For two-body decays {\it only}, we applied an extra factor of
$p^{*2}$ if the decay products were a vector and a pseudoscalar. (Since
this tends to reduce the phase space effect it is clearly conservative).
The widths of the two purely leptonic decay modes of the
$D_s^+$ were added to the sum (a 1.2\%\ correction). Finally, we ignored
the $K^0\overline{K}^0$ and the $K^0_SK^0_SK^0_SK^0_S$ decay modes of
the $D^0$ since the analogues in $D^+_s$ decay are hard to identify. Again, these
decay modes contribute well under 1\%\ of the total $D^0$ width.

      The result of this exercise is shown at the bottom of table
\ref{tbl_d0ds}. The $D^0$ width is larger than the estimated $D^+_s$ width,
predicting that the $D_s^+$ lifetime will be longer than the $D^0$
lifetime by about 25\%. This is to be compared to the measured
difference which is about 17\%. Thus it may well be that the complete
$D_s^+-D^0$ lifetime difference is due merely to phase space effects and
there is no need to invoke any appreciable additional annihilation
contribution to the $D^0$ inclusive decay rate.

\newpage
\begin{center}
\tablecaption{In this table we list the phase space for $D^0$ and corresponding $D^+_s$ decays, 
         the $D^0$ width from the particle tables and finally, the estimated $D^+_s$ width.}

\tablefirsthead {$D^0$ Decay & $D^+_s$ Decay & $D^0$ Phase & $D^+_s$ Phase & $D^0$       & $D^+_s$ Estimated \\ 
       Mode &          Mode &       Space &         Space & Partial Width & Partial Width \\ 
            &               &             &               & ($ps^{-1}$) &  ($ps^{-1}$)  \\ \hline
\vspace*{-10pt} & \\}

\vspace*{3pt}

\begin{supertabular*}{\textwidth} 
 {c@{\extracolsep{\fill}}c@{\extracolsep{\fill}}c@{\extracolsep{\fill}}c@{\extracolsep{\fill}}c@{\extracolsep{\fill}}c@{\extracolsep{\fill}}}
$\mu^+ \nu_\mu                         $  & $\mu^+ \nu_\mu              $             & 1.56575 & 1.56627 & 0.00000   & 0.00927  \\
$\tau^+ \nu_\tau                       $  & $\tau^+ \nu_\tau            $             & 0.14393 & 0.29084 & 0.00000   & 0.01411  \\
$K^-   e^+ \nu_e                       $  & $\eta  e^+ \nu_e
$             & 1.96132 & 1.42768 & 0.08822   & 0.06422  \\
$K^- \mu^+ \nu_\mu                     $  & $\eta \mu^+ \nu_\mu          $            & 1.91829 & 1.39347 & 0.07804   & 0.05669  \\
$K^- \pi^0  e^+ \nu_e     (NR)         $  & $\eta \pi^0  e^+ \nu_e        $           & 0.87300 & 0.58321 & 0.02254   & 0.01506  \\
$\overline{K}^0 \pi^-  e^+ \nu_e (NR)  $  & $\overline{K}^0 K^0  e^+ \nu_e       $    & 0.85393 & 0.45099 & 0.03514   & 0.01856  \\
$K^{*-} e^+ \nu_e                      $  & $\phi  e^+ \nu_e            $             & 0.74673 & 0.66038 & 0.04896   & 0.04330  \\
$\pi^-  e^+ \nu_e                      $  & $K^0  e^+ \nu_e             $             & 3.59061 & 2.28553 & 0.00897   & 0.00571  \\
$K^- \pi^0 \mu^+ \nu_\mu   (NR)        $  & $\eta \pi^0 \mu^+ \nu_\mu
$          & 0.75653 & 0.50155 & 0.02254   & 0.01494  \\
$\overline{K}^0 \pi^- \mu^+ \nu_\mu (NR)$ & $\overline{K}^0 K^0 \mu^+ \nu_\mu     $   & 0.73938 & 0.38250 & 0.03514   & 0.01818  \\
$K^{*-} \mu^+ \nu_\mu                  $  & $\phi \mu^+ \nu_\mu          $            & 0.71949 & 0.63552 & 0.04896   & 0.04324  \\
$\pi^- \mu^+ \nu_\mu                   $  & $K^0 \mu^+ \nu_\mu           $            & 3.51730 & 2.24131 & 0.00897   & 0.00571  \\
$K^- \pi^+                             $  & $\eta \pi^+                $              & 1.45053 & 1.31153 & 0.09283   & 0.08393  \\
$\overline{K}^0 \pi^0                  $  & $K^+ \overline{K}^0               $       & 1.44938 & 1.35708 & 0.05114   & 0.04788  \\
$\overline{K}^0 \pi^+ \pi^-      (NR)  $  & $K^+ K^- \pi^+          (NR)$             & 1.67444 & 1.31914 & 0.03563   & 0.02807  \\
$K^- \pi^+ \pi^0         (NR)          $  & $K^- K^+ \pi^+              $             & 1.69923 & 1.31914 & 0.01672   & 0.01298  \\
$\overline{K}^0 \pi^0 \pi^0      (NR)  $  & $\overline{K}^0 K^+ \pi^0           $     & 1.68900 & 1.31547 & 0.01890   & 0.01472  \\
$K^- \pi^+ \pi^+ \pi^-     (NR)        $  & $K^- K^0 \pi^+ \pi^+          $           & 0.53702 & 0.24755 & 0.04217   & 0.01944  \\
$\overline{K}^0 \pi^+ \pi^- \pi^0  (NR)$  & $\overline{K}^0 K^+ \pi^+ \pi^-       $   & 0.53483 & 0.24755 & 0.05090   & 0.02356  \\
$K^- \pi^+ \pi^0 \pi^0                 $  & $K^- K^+ \pi^0 \pi^+          $           & 0.55103 & 0.25678 & 0.36355   & 0.16941  \\
$K^- \pi^+ \pi^+ \pi^- \pi^0 (NR)      $  & $K^- K^0 \pi^+ \pi^+ \pi^0      $         & 0.05758 & 0.01268 & 0.00242   & 0.00053  \\
$\overline{K}^0 \pi^+ \pi^+ \pi^- \pi^-$  & $\overline{K}^0 K^+ \pi^+ \pi^- \pi^0   $ & 0.05473 & 0.01268 & 0.01406   & 0.00326  \\
$\overline{K}^0 \pi^+ \pi^- \pi^0 \pi^0$  & $\overline{K}^0 K^+ \pi^0 \pi^0 \pi^0   $ & 0.05734 & 0.01354 & 0.25691   & 0.06068  \\
$\overline{K}^0 K^+ K^-        (NR)    $  & $\eta \overline{K}^0 K^+           $      & 0.31535 & 0.19602 & 0.01236   & 0.00768  \\
$K^- K^+ K^- \pi^+                     $  & $\eta K^+ K^- \pi^+          $            & 0.00769 & 0.00726 & 0.00051   & 0.00048  \\
$K^- K^+ \overline{K}^0 \pi^0          $  & $\eta K^+ \overline{K}^0 \pi^0       $    & 0.00769 & 0.00725 & 0.01745   & 0.01646  \\
$\overline{K}^0 \eta                   $  & $\overline{K}^0 K^+               $       & 1.30045 & 1.35708 & 0.01697   & 0.01770  \\
$\overline{K}^0 \rho^0                 $  & $K^{*+} \overline{K}^0            $       & 1.13230 & 1.08952 & 0.02933   & 0.02912  \\
$K^- \rho^+                            $  & $\eta \rho^+$
& 1.13507 & 0.93361 & 0.26175   & 0.16237  \\
$\overline{K}^0 \omega                 $  & $\overline{K}^0 K^{*+}$                   & 1.12856 & 1.08952 & 0.05090   & 0.05105  \\
$\overline{K}^0 \eta^{\prime-}(958)    $  & $\overline{K}^0 K^+               $       & 0.95141 & 1.35708 & 0.04144   & 0.05912  \\
$\overline{K}^0 f_0(980)               $  & $\overline{K}^0 K^{*0}(1430)^+    $       & 0.92292 & 0.33782 & 0.01381   & 0.00506  \\
$\overline{K}^0 \phi                   $  & $\overline{K}^0 K^{*+}            $       & 0.87666 & 1.08952 & 0.02084   & 0.04460  \\
$K^- a1(1260)                          $  & $\overline{K}^0 K_{1A}^+          $       & 0.51031 & 0.47496 & 0.17693   & 0.15902  \\
$\overline{K}^0 f_2(1270)              $  & $\overline{K}^0 K_2^*(1430)^+     $       & 0.40873 & 0.27454 & 0.00994   & 0.00667  \\
$\overline{K}^0 f_0(1370)              $  & $\overline{K}^0 K^*_0(1430)^+     $       & 0.28812 & 0.33782 & 0.01672   & 0.01961  \\
$K^{*-} \pi^+                          $  & $\phi \pi^+   $                           & 1.19686 & 1.13584 & 0.12118   & 0.11546  \\
$\overline{K}^{*0} \pi^0               $  & $\overline{K}^{*0} K^+          $         & 1.19413 & 1.08819 & 0.07513   & 0.06338  \\
$\overline{K}^{*0} \pi^+ \pi^-         $  & $\phi \pi^+ \pi^0        (NR)$            & 0.55761 & 0.49927 & 0.05332   & 0.04774  \\
$K^- \pi^+ \rho^0                      $  & $\eta \pi^+ \rho^0        $               & 0.27779 & 0.17306 & 0.15269   & 0.09513  \\
$K^{*-} \rho^+                         $  & $\phi \rho^+              $               & 0.67368 & 0.61124 & 0.14784   & 0.13414  \\
$K1(1270)^- \pi^+                      $  & $f_1(1285) \pi^+          $               & 0.81186 & 0.88490 & 0.02569   & 0.03709  \\
$K^*_0(1430)^- \pi^+                   $  & $f_0(980) \pi^+           $               & 0.63668 & 1.16733 & 0.02521   & 0.04621  \\
$\overline{K}^{*0} \eta                $  & $\overline{K}^{*0} K^+          $         & 0.97557 & 1.08819 & 0.04605   & 0.07125  \\
$K^- \pi^+ \omega      (NR)            $  & $\eta \pi^+ \omega         $              & 0.23732 & 0.14638 & 0.04605   & 0.02840  \\
$\overline{K}^{*0} \omega              $  & $K^{*+} \overline{K}^{*0}       $         & 0.67860 & 0.65378 & 0.02666   & 0.02569  \\
$K^- \pi^+ \eta^{\prime-}              $  & $\eta \pi^+ \eta^{\prime-}   $            & 0.06989 & 0.04908 & 0.01697   & 0.01192  \\
$\pi^+ \pi^-                           $  & $\pi^+ K^0                 $              & 1.55309 & 1.46143 & 0.00368   & 0.00347  \\
$\pi^0 \pi^0                           $  & $\pi^0 K^+                 $              & 1.55424 & 1.46363 & 0.00204   & 0.00192  \\
$\pi^+ \pi^- \pi^0                     $  & $\pi^+ K^+ \pi^-             $            & 3.23927 & 2.01929 & 0.03878   & 0.02417  \\
$\pi^+ \pi^- \pi^+ \pi^-               $  & $\pi^+ K^+ \pi^0 \pi^-         $          & 1.87461 & 0.79452 & 0.01769   & 0.00750  \\
$\pi^+ \pi^- \pi^+ \pi^- \pi^0         $  & $\pi^+ K^+ \pi^+ \pi^- \pi^-     $        & 0.41812 & 0.10642 & 0.04605   & 0.01172  \\
$\pi^+ \pi^- \pi^+ \pi^- \pi^+ \pi^-   $  & $\pi^+ K^+ \pi^+ \pi^- \pi^- \pi^0 $      & 0.03757 & 0.00537 & 0.00097   & 0.00014  \\
$K^+ K^-                               $  & $\eta K^+                 $               & 1.33247 & 1.18007 & 0.01030   & 0.00912  \\
$K^0 K^- \pi^+          (NR)           $  & $\eta K^0 \pi^+             $             & 1.01935 & 0.68613 & 0.00557   & 0.00375  \\
$\overline{K}^0 K^+ \pi^-       (NR)   $  & $K^0 \overline{K}^0 K^+            $      & 1.01935 & 0.49496 & 0.00921   & 0.00447  \\
$K^+ K^- \pi^0                         $  & $\pi^+ \pi^- \pi^+            $           & 1.03921 & 3.67230 & 0.03151   & 0.11134  \\
$K^+ K^- \rho^0                        $  & $K^+ K^- K^{*+}            $              & 0.04758 & 0.01903 & 0.00218   & 0.00087  \\
$K^0 \overline{K}^0 \pi^+ \pi^-        $  & $\pi^+ K^0 \overline{K}^0 K^0        $    & 0.13646 & 0.01137 & 0.01648   & 0.00137  \\
$K^+ K^- \pi^+ \pi^- \pi^0             $  & $K^+ K^- \pi^+ \pi^- K^+       $          & 0.00464 & 0.00009 & 0.00751   & 0.00015  \\
$K^{*+} K^-                            $  & $\eta K^{*+}              $               & 1.02605 & 0.78911 & 0.00848   & 0.00430  \\
$K^{*-} K^+                            $  & $\phi K^+                 $               & 1.02605 & 0.96836 & 0.00436   & 0.00409  \\
$\phi \rho^0         (NR)              $  & $\phi K^{*+}              $               & 0.39711 & 0.35351 & 0.00145   & 0.00129  \\
$\phi \pi^+ \pi^-        (NR)          $  & $\phi K^+ \pi^0             $             & 0.34522 & 0.17350 & 0.00170   & 0.00085  \\
$\overline{K}^{*0} K^{*0}              $  & $\phi K^{*+}            $                 & 0.41460 & 0.35351 & 0.00339   & 0.00289  \\
\hline
Total                                     & Total                                     & \omit   & \omit   & 2.95982   & 2.22225  \\
\hline
\end{supertabular*}
\label{tbl_d0ds}
\end{center}

\newpage
\section{Reduced $D_s^+$ inclusive decay rate via a reduced effective
         charm quark mass} 
 
      Phase space effects manifest also via the
effective mass of the decaying $c$ quark. The binding of the $c$ quark
is stronger in the $c\overline{s}$ system then in the $c\overline{u}$ 
system. Since the fifth power of this mass enters in $\Gamma(Q\to
ql\nu)$ small shifts of $m(Q)$ may have enhanced, perceptible effects in
the spectator model.
This is analogous \cite{ros} to the reduction due to 
the Coulombic binding $(Z\alpha^2/2)m_\mu$ of the $\mu^-\to
e^-\nu_\mu\overline{\nu}_e$ decay rate in muonic atoms.
The asymptotic physical state after the $\mu^-$ decay no longer has 
Coulomb binding. Consequently the effective mass of the decaying muon and the 
corresponding phase space and decay rate will be reduced by
$1-r = (1- \hbox{BE}/m_\mu)$ and by $(1-r)^5$ respectively, where
$\hbox{BE}$ is the binding energy of the muon.

The present case is not as clear cut. 
Due to confinement the spectator 
quark of the initial $D$ ``bound state''
and the three quarks from the decay of the $c$ quark 
appear (albeit in new rearranged forms) as constituents of 
the final hadrons. 

Thus color binding is not completely lost here. This is in
contradistinction 
with the loss of Coulombic binding in the above $\mu^-$ decay.
Rather we have here only an enhanced binding of the $c\overline{s}$ (or
$D_s^+$) system as 
compared with the $c\overline{u}$ or $D^0$ system. 
This ``relatively'' low $m_{D_s^+}$ mass is indeed the source of the phase 
space reduction in the actual physical decay channels of the $D_s^+$ 
relative to that of the $D^0$ which is precisely what has been discussed
in section II.
Hence, whatever ``effect'' is eventually found in the present section, does not 
add up to the above suppression. Rather, it is just another way, 
utilizing quark hadron duality, of restating the same effect or 
part thereof.
In the spirit of the spectator model of the decay we consider 
only the reduction due to tighter binding of the $c$ quark mass in 
$D_s^+ = c\overline{s}$ 
as compared with that of the $c$ quark inside $D^0 = c\overline{u}$. Since binding 
effects are often deemed to be pertinent only to the interacting two body 
system as a whole i.e., to the $c$ {\it and} $\overline{q} = \overline{s}$ (or
$\overline{u}$) the separation of the 
``charm quark associated part'' is, at best, ambiguous.
 
      If the $Q\overline{q}$ system was naively treated as a
non-relativistic Coulomb two-body system then the binding energy $BE =
{\alpha_S^2\over 2}m_{Q, q}$ where the reduced mass $m_{Q, q}$ is given by

\begin{eqnarray}
  m_{Q,q} &=& {m_{q}m(Q)\over m_{Q}+ m_{q}}
\label{eqn_redmass}
\end{eqnarray}

Using $m_{c} = 1.5$ GeV, $m_{u} = 0.3$ GeV and $m_{s} = 0.45$ GeV, we find that the ratio of 
reduced masses in the $D_s^+$ and $D^0$ system is about 0.8
and their difference is of order 100 MeV. The Coulombic binding
difference between $D_s^+$ and $D^0$ is then $BE = (\alpha_S^2 / 2) 100$ MeV = 50 MeV,
where we arbitrarily used $ \alpha_S = 1$. If half of this is assigned to the 
charmed quark then its effective mass in the $D_s^+$ decay will be 25/1500
= 1.6\%\ lower than that in $D^0$ decays. The resulting phase space reduction due to  
$m_c^5$ is then 8\%. The Coulombic interaction maximizes the 
effect of quark (reduced) mass changes, as in this case the reduced
quark mass $m_{Q,q}$ is the 
only relevant parameter of mass dimension entering the problem.
Once other-dimensional parameters 
are introduced this dependence on $m_{Q,q}$ gets diluted and weakened. In 
particular this is the case for any effective interaction 

\begin{eqnarray}
  V(r) &=& r^\beta \qquad\qquad \hbox{with} \qquad\qquad 1 > \beta > -1
\label{eqn_Vr}
\end{eqnarray}
 
      Such potentials may indeed be more appropriate to describe the
$Q\overline{q}$ interaction at the $\sim 0.5$ fm
scale of the physical bound state. Thus the spectator model phase
space effect is smaller than 8\%. It is therefore unlikely to account,
by itself, for the observed 17\%\ reduction in $\Gamma(D_s^+)$ when
compared to $\Gamma(D^0)$.
  
\section{Multiple pion states in the decay of the tau lepton}

Upper bounds on the mass of the tau neutrino were deduced from 
analysis of the decay $\tau^-\to\pi^+\pi^+\pi^-\pi^-\pi^-\nu_\tau$.
The seven pion mode,
$\tau^-\to\pi^+\pi^+\pi^+\pi^-\pi^-\pi^-\pi^-\nu_\tau$ if appreciable, could lead to 
a more stringent bound. The OPAL and CLEO collaborations searched for 
these decays. Finding none they deduced the following upper bounds

\begin{eqnarray}
  BR(\tau^-\to\pi^+\pi^+\pi^+\pi^-\pi^-\pi^-\pi^-\nu_\tau) &<& 1.8
\times 10^{-5} \qquad \hbox{(OPAL)}
\label{eqn_OPAL}
\end{eqnarray}
and 
\begin{eqnarray}
  BR(\tau^-\to\pi^+\pi^+\pi^+\pi^-\pi^-\pi^-\pi^-\nu_\tau) &<& 2.6
\times 10^{-6} \qquad \hbox{(CLEO)}
\label{eqn_CLEO}
\end{eqnarray}

at the 95\%\ CL.

      These rare tau decays are intersting in a purely hadronic 
context as well. Many ``soft'' pions are produced with 
small (L=0 or L=1) overall angular momenta. These decays may provide a 
simpler setting than $p\overline{p}$ annihilations at rest for studying such 
systems. Here, but not in the annihilation, the pions are produced via the 
local vector/axial currents. This in particular allows us to estimate 
the expected rates via chiral Lagrangians \cite{wess}. We find, 
mainly due to  phase space limitations, that the expected rate is much 
smaller than the existing bounds. Thus finding even ONE unambiguous 
such decay would indicate strong ``Dynamical'' enhancements. The latter 
could in particular indicate the onset of non-perturbative
``Condensate'' effects operative only for a sufficiently large number of
pions. 

In order to estimate the decay rate
$\Gamma(\tau^-\to\pi^+\pi^+\pi^+\pi^-\pi^-\pi^-\pi^-\nu_\tau)$ we
utilize the simplest chiral Lagrangian

\begin{eqnarray}
  {\cal L} &=& 2f_\pi^2(\exp({i\hat\phi\over 2f_\pi})\lrp_\mu\exp({-i\hat\phi\over 2f_\pi}))^2
\label{eqn_lrp1}
\end{eqnarray}

with 

\begin{eqnarray}
  \hat\phi &=& \phi_i\tau_i
\label{eqn_phihat}
\end{eqnarray}

and $f_\pi \approx 90$ MeV.

The corresponding Noether current:

\begin{eqnarray}
  {\cal J}_\mu &=& {\partial\over\partial_\mu\phi}{\cal L} = 
  2f_\pi^2(\exp({i\hat\phi\over 2f_\pi})\lrp_\mu\exp({-i\hat\phi\over 2f_\pi}))
\label{eqn_lrp2}
\end{eqnarray}
  
yields the vector / axial currents as its $\hat\phi$ even / odd
parts. We can, as indicated in Appendix I, extract the seventh order
term 
\begin{eqnarray}
  c_7(\partial_\mu\hat\phi){(\hat\phi\cdot\hat\phi)^3\over (2f_\pi)^5}
\label{eqn_c7val}
\end{eqnarray}

relevant to the 
$\bra{0}J^A_\mu(x)\ket{\pi_1\pi_2\ldots\pi_7}$
amplitude of interest. 

Had we started from a more elaborate Lagrangian we might have generated 
$J_A^\mu(x)$ with extra derivatives. However in the ``Soft 
pion'' limit - which is almost achieved here - the minimal form of
equation [\ref{eqn_c7val}] with just one derivative may dominate.
  
      The complete effective Lagrangian for the tau decay also contains the 
lepton current:

\begin{eqnarray}
  {\cal L} &=& {G_F\over\sqrt{2}}\int\,d^4x
\overline{\Psi}_\tau(x)\gamma_\mu\gamma_5\Psi_\nu(x)c_7(\partial_\mu\phi)\phi^6
\label{eqn_taulag}
\end{eqnarray}

using $\partial_\mu\phi^7 \approx 7(\partial_\mu\phi)\phi^6$, 
$\parslash\Psi_\tau(x) = m_\tau\Psi_\tau(x)$, $\parslash\Psi_\nu(x)=0$,
$\gamma_5\Psi_\nu = \Psi_\nu$ and 
intergrating by parts we can rewrite the decay Lagrangian as:

\begin{eqnarray}
  {\cal L}^{\hbox{eff}} &=& {c_7\over 7}{G_f\over(2f_\pi)^5}m_\tau\Psi_\tau(x)\Psi_{\nu_\tau}(x)\phi^7(x)
\label{eqn_efflag}
\end{eqnarray}

      In momentum space $\overline{\Psi}_\tau\gamma_5\Psi_\nu$ becomes
$\overline{u}_\tau(p)u_\nu(q)$ with $p$, $q$ the four momenta of 
the tau and tau neutrino. Upon squaring and spin summation this yields an extra
$m(\tau)$ factor (the massless tau neutrino is conventionally normalized
to $\overline{u}u=1$). Thus the Lagrangian (\ref{eqn_efflag}) yields:

\begin{eqnarray}
  \Gamma(\tau^-\to\pi^+\pi^+\pi^+\pi^-\pi^-\pi^-\pi^-\nu_\tau) &=& 
  \left(c_7\over 7\right)^2 {G_F^2 m_\tau^3 
    {\overline\Phi}(m_\tau, m_\pi, m_\pi, m_\pi, m_\pi, m_\pi, m_\pi,
      m_\pi, 0)\over (2f_\pi)^{10} (2\pi)^{24}}
\label{eqn_gam7pi}
\end{eqnarray}
 
with $\overline\Phi$ being the eight body phase space for the 
$\tau^-\to\pi^+\pi^+\pi^+\pi^-\pi^-\pi^-\pi^-\nu_\tau$ decay.

The above phase space and for the other tau decays into states with $n$ 
final pions are evaluated via the composition formula

\begin{eqnarray}
  \Phi_{(n_1 + n_2)}(W) &=&
    \prod\int_{(n_1m_\pi)^2}^\infty\,dW_1^2\;\int_{(n_2m_\pi)^2}^\infty\,dW_2^2 
      {\Delta^{1/2}(W, W_1, W_2)\over
       2W^2}\Phi_{n_1}(W_1)\Phi_{n_2}(W_2) \Theta(\Delta) 
\label{eqn_psp}
\end{eqnarray}

where 

\begin{eqnarray}
  \Delta(W, W_1, W_2) &=& [W^2 - (W_1 + W_2)^2] [W^2 - (W_1 - W_2)^2]
\label{eqn_heron}
\end{eqnarray}

is the ``Triangle Function'' (Heron's expression for the square of the area of the triangle in terms 
of the lengths of its sides) of W, the invariant mass of the complete $n$ 
particle system and $W_1$, $W_2$ are the masses of the subsystems with $n_1$ , $n_2$  
particles. $\Theta(x)$ is the unit step function. The resulting phase space values for 
$n = 1, 2, \ldots 8$ are presented in table \ref{tbl_phsp} below. For masses,
energies and momenta we use the GeV unit throughout this paper. In
particular, the dimensions of the LIPS for $n+1$ particles is
GeV$^{(2n-2)}$. 
Following the convention of 
Perl\cite{perl} we do not include the $(2\pi)^{-3}$ normalization
factors for each final state particle in the phase space. 

\begin{table}[htbp]
\caption{In this table we list the phase space for $\tau$ decays to a
         neutrino and $n$ charged pions.}
\vspace*{3pt}
\begin{tabular}{cccccc}\hline
$n$ in $\tau \to n\pi$ & Phase Space \\ 
1 &      1.56111 \\
2 &      3.10573 \\
3 &      1.75786 \\
4 &     0.383552 \\
5 &    0.0354896 \\
6 &   0.00139612 \\
7 &  2.17774$\times 10^{-5}$ \\
8 &  1.15161$\times 10^{-7}$ \\
\hline
\end{tabular}
\label{tbl_phsp}
\end{table}

      To estimate
$\Gamma(\tau^-\to\pi^+\pi^+\pi^+\pi^-\pi^-\pi^-\pi^-\nu_\tau)$ 
we compare it with the decay into $\nu_\tau$ and five charged pions with a 
measured BR $\sim 1$\%\ and a theoretical estimate analogous to
equation \ref{eqn_gam7pi}:

\begin{eqnarray}
  \Gamma(\tau^-\to\pi^+\pi^+\pi^-\pi^-\pi^-\nu_\tau) &=& 
  \left(c_5\over 5\right)^2 {G_F^2 m_\tau^3 
    {\overline\Phi}(m_\tau, m_\pi, m_\pi, m_\pi, m_\pi, m_\pi, 0) \over (2f_\pi)^{6} (2\pi)^{18}}
\label{eqn_gam5pi}
\end{eqnarray}

using equations \ref{eqn_gam7pi} and \ref{eqn_gam5pi}:

\begin{eqnarray}
  {BR(\tau^-\to\pi^+\pi^+\pi^+\pi^-\pi^-\pi^-\pi^-\nu_\tau) \over
   BR(\tau^-\to\pi^+\pi^+\pi^-\pi^-\pi^-\nu_\tau)}
   &=& 
  {{\overline\Phi}(m_\tau, m_\pi, m_\pi, m_\pi, m_\pi, m_\pi, m_\pi, m_\pi, 0)\over
   (2\pi)^6(2f_\pi)^4{\overline\Phi}(m_\tau, m_\pi, m_\pi, m_\pi, m_\pi, m_\pi, 0)}
  \left(5c_7\over 7c_5\right)^2\\
   &=&
  6\times 10^{-6}\left(5c_7\over 7c_5\right)^2
\label{eqn_taurate}
\end{eqnarray}

      Using the 1\%\ branching ratio into 5 charged pions we then have 

\begin{eqnarray}
  BR(\tau^-\to\pi^+\pi^+\pi^+\pi^-\pi^-\pi^-\pi^-\nu_\tau) &=&
  6\times 10^{-8}\left(5c_7\over 7c_5\right)^2
\label{eqn_7piBR2}
\end{eqnarray}

      In Appendix I we find that $(5c_7 / 7c_5) = 1/30$ and thus
equation \ref{eqn_7piBR2} leads to a hopelessly small branching ratio 
$BR(\tau^-\to\pi^+\pi^+\pi^+\pi^-\pi^-\pi^-\pi^-\nu_\tau) = 6\times
10^{-11}$.\cite{fn2} 

      The chiral Lagrangian approach used here is justified {\bf all} 
momenta in the problem are small. For the case 
at hand the total invariant mass of the seven pion hadronic system
and masses of 3, 4, and 5 pion subsystems are typically high ($W\approx 1$ GeV). 
Various resonances $a_1(1240)$  $a_2(1320)$, $\pi(1300)$ etc. can then appear in 
these subchannels. These resonances could significantly enhance certain 
rates and invalidate the local effective Lagrangian approach.

      We would like however to argue that while resonances typically
dominate decays of the tau lepton (and also of charmed hadrons) into 2-4 particles,
their effect on the multi-particle decays of interest with many soft slow
pions may be minimal.  The point is that these heavy resonances with
prominent multiparticle decays are very broad, $\Gamma = 300 - 400$
MeV. These resonances also move rather slowly ($\beta < 1/2$) and the
decay products of different resonances emerge within a common
interaction range. Hence resonance (re-)formations and decays can, in
the context of the present final state of many, highly overlapping, soft
pions be subsumed into the local effective Lagrangian. 

In the momentum
space description the $n$-particle vertex is effectively local so long
as the {\it real} off-shellness of the internal particles exchanged OR
the {\it imaginary} part therof (viz. widths) are large in comparison
with the final physical particle momenta. 

The effect of
resonances is likely to be stronger when fewer particles occur in the
final state. Hence our value for the 7$\pi$ branching ratio obtained by treating
both this and the 5$\pi$ decay by the same chiral Lagrangian and not
accounting for possible resonance enhancement of the latter may well
yield an overestimate of $BR(\tau\to 7\pi\nu_\tau)$.

      The situation is drastically different when the multipion final
state is dominated by {\it narrow} resonances. This, in particular, is
the case for the $\tau\to 6\pi\nu_\tau$ recently observed by
CLEO.\cite{cleo} These observations are consistent with being of the form
$\tau\to \eta 3\pi \nu_tau$ and $\tau\to \omega 3pi\nu_tau$ with
subsequent decays of the $\eta$ and $\omega$ to $3\pi$. 

      Low energy $\pi\pi$ scattering in general and scattering lengths
in particular are studied in simpler settings such as $K_{l4}$ decays or
the $\pi N \to \pi\pi N$ reaction in a region dominated by an almost
real pion exchange. The resulting S-wave scattering lengths
          
\begin{eqnarray}
  a(I&=&0) \approx 0.22/m_\pi \qquad\qquad \hbox{and} \qquad\qquad a(I=2) \approx
-(0.02 - 0.08)/m_\pi
\label{eqn_avals}
\end{eqnarray}

are relatively small in agreement with various theoretical 
considerations \cite{donoghue}, \cite{leutwyler}.

      The many-pion final states of interest could 
be particularly sensitive to the threshold $\pi\pi$ interaction. To
analyze this we use
a non-relativistic description. The total 
interaction energy in the $n$ pion state, when all pions are within 
interaction range, increases {\it quadratically} 
with $n$:

\begin{eqnarray}
  U &=& \sum V_{ij} (r_i - r_j) \approx n(n-1)/2 <V>
\label{eqn_U}
\end{eqnarray}
where $<V>$ indicates the expectation value of an attractive
interaction. 

As we shortly indicate, this interaction is attractive.
If the pions can be treated as pointlike elementary bosons this will 
always lead eventually to condensation. All pions 
can then be put into a common state - the (S-wave) ground state 
inside a spherical cavity of radius $r$ smaller than the $\pi$--$\pi$ 
interaction range which is denoted by $r$.  The kinetic energy  
$T = n \hbar^2 / 2mr^2$ (or $n\hbar/r$) which is linear in $n$ is less than the attractive
potential energy of equation \ref{eqn_U}.

\begin{eqnarray}
  n &>& 1 + {2 <T> \over |<V>|}
\label{eqn_nmin}
\end{eqnarray}

  For $n$ larger than $n_c$ defined by
         
\begin{eqnarray}
  n_c &=& 1 + {2 m_\pi\over <V>}
\label{eqn_nc}
\end{eqnarray}

the interaction energy exceeds even the rest mass of the pions.
The description in terms 
of pions ceases then to be useful and the system is better 
described by the $\phi$ (or $\phi^2$) field. In particular the rate for 
producing more than $n_c$ pions in $\tau\to n\pi + \nu_\tau$ could be
dramatically enhanced. 
 
As we will argue next, the requisite $n_c$ is far higher than the number of 
pions in a single cluster produced in tau decays or $\overline{p}p$
annihilations at rest.

Let $n = 2 k + 1$ so that the $\tau^+$ decays into $k+1$ positive and
$k$ negative pions. We have $k(k+1)/2$ $++$ pairs and k(k-1)/2 $--$
pairs, a total of $k^2$ pairs which are in a pure I=2 state, and
$k(k+1)$ $+-$ pairs which constitute I=0 and I=2 states in a 2:1
ratio. (Since all pairs are in relative S-wave states, Bose statistics
forbids any I=1 states). The effective S-wave scattering length averaged
over all $(2k+1)k$ possible pairs is therefore:

\begin{eqnarray}
  a_{\hbox{eff}} &=& {[k^2 + k(k+1)/3] a(I=2) + (2/3) k(k+1) a(I=0)\over (2k+1)k}\\
    &=& {(4k+1) a(I=2) + (2k+2) a(I=0)\over 3(2k+1)}
\label{eqn_aeff}
\end{eqnarray}

For large $k$ we simply have

\begin{eqnarray}
  a_{\hbox{eff}} &=& {2 a(I=2)\over 3} + {a(I=0)\over 3} \approx (0.06 - 0.02) m_\pi^{-1}
\label{eqn_aeff2}
\end{eqnarray}

and in particular, as stated above, is attractive.

      Let us next proceed to a rough estimate of the critical number of particles 
which is the threshold for the onset of condensation from either
equation \ref{eqn_nmin} or equation \ref{eqn_nc} above. 
For simplicity the $\pi$ - $\pi$ interaction is approximated by a square
well of depth $-V_0$ and radius $r_0$.
Elementary considerations then yield $a = \tan(qr_0)/q - r_0$, where
$q=\sqrt{m_\pi V_0}$. 

Since there is no near threshold $\pi-\pi$ bound state and $qr_0 \le 1$ 
we expand $\tan(qr_0)$, finding 

\begin{eqnarray}
  a \approx {(qr_0)^3\over 3q} &=& {1\over 3} m_\pi V_0r_0^3
\label{eqn_qr01}
\end{eqnarray}

We next used equation \ref{eqn_qr01} and estimates of the expectation
values $<T>$ and $<V>$ pertinent to this specific square well model to
obtain from equation \ref{eqn_nmin}: 

\begin{eqnarray}
  n - 1 > {(\pi)^2 r_0 \over 3a} &=& 16 - 48
\label{eqn_qr02}
\end{eqnarray}

The actual numbers on the right hand side of equation \ref{eqn_qr02}
were obtained for $r_0 = 1/(3m_\pi)$ and the range of $a_{eff}$ in
equation \ref{eqn_aeff2}. (We note that since for this $r_0$ value 
$<T> \approx 1/r_0 \ge m_\pi$ equation \ref{eqn_nmin} implies equation
\ref{eqn_nc}). 

Even if we use $r_0 = 1/(5m_\pi)$ corresponding to an exchange of a ``$\sigma$'' particle of 
mass 700 MeV, $n$ is still in the range 10 -- 30.

      The above considerations strongly suggest that for the 7-pion final 
state of interest any condensate enhancements are most unlikely and our 
above estimates of rates should stand.

\section{Acknowledgements}

      We would like to thank Silas Beans for his help and advice
regarding the chiral Lagrangian calculations presented in the Appendix
and to Manoj Banerjee for his interest. S.~N. thanks the Israeli
National Science Foundation (grant 561/99) and M.~V.~P. thanks the
U.S. Department of Energy.

\section{Appendix I}

We will estimate the decay rate 
$\Gamma (\tau^- \rightarrow \pi^+\pi^+\pi^+\pi^-\pi^-\pi^-\pi^-\nu_{\tau})$ using
the leading order axial vector contribution in chiral perturbation theory.
The leading order chiral Lagrangian is given by

\begin{equation}
{\cal L}={1\over 4}\, f_\pi^2\, Tr( \partial_\mu \Sigma^\dagger \partial_\mu \Sigma ).
\label{lagr}
\end{equation}

The field $\Sigma$ transforms linearly with respect to $SU(2)_L\times
SU(2)_R$: $\Sigma\rightarrow {L} \Sigma {R^\dagger}$ where ${{L,R}}$
is an element of $SU(2)_{L,R}$.  A convenient parameterization of
$\Sigma$ is

\begin{equation}
\Sigma =\exp({{i \phi_a \tau_a}\over{f_\pi}})
\end{equation}
where $f_\pi =93$ MeV and the $\tau_a$ are the Pauli matrices.  The
Noether current associated with $SU(2)_{L}$ is

\begin{equation}
J^\mu_{La} =i{1\over 4}\, f_\pi^2\, Tr( \Sigma^\dagger \tau_a \partial^\mu \Sigma ).
\end{equation}
With the help of a useful mathematical formula~\cite{georgi} one finds

\begin{equation}
J^\mu_{La} =-{1\over 4}\, f_\pi\, \int_0^1 ds\,
Tr \left[ \tau_a \left( \partial^\mu {\tilde\phi} 
+ {{is}\over{f_\pi}} [{\tilde\phi},\partial^\mu {\tilde\phi}] 
+ {1\over 2!}\left({{is}\over{f_\pi}}\right)^2 
[{\tilde\phi},[{\tilde\phi},\partial^\mu {\tilde\phi}]] 
+{\cal O}(s^3 )\right)\right]
\end{equation}
where ${\tilde\phi}\equiv \phi_a\tau_a$.  The $SU(2)_{R}$ current can then be
obtained by switching the signs of all the odd terms in the pion
field. The axial vector current is 

\begin{equation}
A^\mu_{a}
 =-{1\over 2}\, f_\pi\, \int_0^1 ds\,
Tr \left[ \tau_a \left( \partial^\mu {\tilde\phi} + 
{1\over 2!}\left({{is}\over{f_\pi}}\right)^2 
[{\tilde\phi},[{\tilde\phi},\partial^\mu {\tilde\phi}]] 
+{\cal O}(s^4 )\right)\right].
\end{equation}
It is then straightforward to find

\begin{equation}
A^\mu_{a}(n)
 =-{1\over 2}\, f_\pi\, {1\over{n!}}\left({i\over{f_\pi}}\right)^{n-1}
Tr \left( \tau_a
[{\tilde\phi},\ldots,[{\tilde\phi},[{\tilde\phi},
[{\tilde\phi},\partial^\mu {\tilde\phi}],\ldots,]]] \right).
\end{equation}
where $n$ is the number (odd) of pions. For instance,

\begin{equation}
A^\mu_{a}(1)
 =-{1\over 2}\, f_\pi\, 
Tr \left( \tau_a \partial^\mu {\tilde\phi} \right)=
-f_\pi\, \partial^\mu {\phi_a};
\end{equation}

\begin{equation}
A^\mu_{a}(3)
 ={1\over{ 12 f_\pi}}
Tr \left( \tau_a
[{\tilde\phi},[{\tilde\phi},\partial^\mu {\tilde\phi}]]
\right);
\end{equation}

\begin{equation}
A^\mu_{a}(5)
 =-{1\over{ 240 f_\pi^3}}
Tr \left( \tau_a
[{\tilde\phi},[{\tilde\phi},[{\tilde\phi},[{\tilde\phi},\partial^\mu {\tilde\phi}]]]] 
\right);
\label{eqA5}
\end{equation}

\begin{equation}
A^\mu_{a}(7)
 ={1\over{10080 f_\pi^5}}
Tr \left( \tau_a
[{\tilde\phi},[{\tilde\phi},[{\tilde\phi},
[{\tilde\phi},[{\tilde\phi},[{\tilde\phi},\partial^\mu {\tilde\phi}]]]]]]
\right).
\end{equation}

Using $[\tau_i, \tau_j] = \epsilon_{ijk}\tau_k$ we can write 

\begin{equation}
A^\mu_{a}(7) = {1\over 10080 f_\pi^5}\epsilon_{abc}\epsilon_{edc}\epsilon_{fge}\epsilon_{ihf}\epsilon_{kji}\epsilon_{mlk}
\phi_l \phi_j \phi_h \phi_g \phi_d \phi_b \partial_\mu \phi_a
\end{equation}

By pairwise contraction of $\epsilon$ symbols we can reduce the
expression into the desired 
$c(7) (\phi\cdot\phi)^3 \partial_\mu\phi$ form.
Likewise we generate from $A_\mu(5)$ of Eq.~(\ref{eqA5}) above an
analogous $c_5(\phi\cdot\phi)^2\partial_\mu\phi$ term.
We find that $c(7) = 2^3/10080$ and $c(5) = 2^2/240$. Hence, in particular, 
$5c(7) / 7c(5) = 10 / 294 \approx 1/30$.

Actually all currents $A(1)$ -- $A(7)$ contribute to the decay $\tau^- \rightarrow
\pi^+\pi^+\pi^+\pi^-\pi^-\pi^-\pi^-\nu_{\tau}$ at leading order in
chiral perturbation theory. For instance, $A^\mu_{a}(5)$ contributes
with one of the pions going to three pions through the $4$-point
interaction contained in Eq.~(\ref{lagr}). In order to evaluate these we
need more involved calculations. Thus the
last contribution is the integral over $\sigma = \sqrt{s(678)}$ of the product
of the $A(5)$ amplitude with one pion off-shell at $\sigma$ and the $\pi\pi$
scattering with one pion off-shell at the same $\sigma$. 
All these
contributions should be comparable and we do not expect these to
drastically change our estimate of the $c(7)/c(5)$ ratio.


\begin{thebibliography}{}
\bibitem{fn1} Note that the spectator and annihilation decays 
correspond (at the quark level) to different two and four quark states. 
The amplitudes for annihilation and direct $Q$ decay interfere in 
specific (hadronic) decay channels, but not in the inclusive
rates. Hence one adds the two contributions in the total decay rate.
\bibitem{fn2} If we were exactly at threshold, with all the pions in a
common S-wave state, the rate is enhanced by a ``Bose-Einstein'' 
factor $n_{\pi^+}!n_{\pi^-}!$. Since the pions in the 7$\pi$ decay
are slower than those in the 5$\pi$ decay the enhancement factor there
is closer to the expected value of $4!\times 3!  = 144$ than the one in
the 5 pion decay, $3!\times 2! = 24$. However, even in the unlikely case
where this enhancement is fully operative in $\tau\to 7\pi\nu_\tau$ and
not at all in $\tau\to 5\pi\nu_\tau$, the 7 pion $BR(\tau\to
7\pi\nu_\tau) = 6\times 10^{-9}$ is still $\sim$400 times smaller than
the present upper bound.
\end{thebibliography}
\end{document}